\documentclass[prl,a4paper,twocolumn,showpacs,groupedaddress]{revtex4-1}

\usepackage{graphicx}
\usepackage{amsmath}
\usepackage{epstopdf}

\usepackage{indentfirst}

\begin{document}

\title{Control of charge migration in molecules by ultrashort laser pulses}

\author{Nikolay V. \surname{Golubev}}
\email[e-mail: ]{nikolay.golubev@pci.uni-heidelberg.de}
\affiliation{Theoretische Chemie, Universit\"at Heidelberg, Im Neuenheimer Feld 229, 69120 Heidelberg, Germany}

\author{Alexander I. \surname{Kuleff}}
\email[e-mail: ]{alexander.kuleff@pci.uni-heidelberg.de}
\affiliation{Theoretische Chemie, Universit\"at Heidelberg, Im Neuenheimer Feld 229, 69120 Heidelberg, Germany}

\date{\today}

\begin{abstract}

Due to electronic many-body effects, the ionization of a molecule can trigger ultrafast electron dynamics appearing as a migration of the created hole charge throughout the system. Here we propose a scheme for control of the charge migration dynamics with a single ultrashort laser pulse. We demonstrate by fully \textit{ab initio} calculations on a molecule containing a chromophore and an amine moieties that simple pulses can be used for stopping the charge-migration oscillations and localizing the charge on the desired site of the system. We argue that this control may be used to predetermine the follow-up nuclear rearrangement and thus the molecular reactivity.

\end{abstract}


\maketitle


Since the early days of quantum mechanics, the aim of the scientific efforts has been not only to understand the microscopic world, but also to use the quantum properties for controlling different processes. For example, one may use the quantum interference and the properties of the laser-matter interaction in order to exert control over the chemical reactivity of a molecule, a research field known nowadays as ``femtochemistry'' \cite{Femtochem_book}. In femtochemistry, by specifically tailored femtosecond laser pulses, one tries to steer the motion of the nuclei in the molecule and thus to bring the system into the desired reaction channel.

In recent years, the rapid development of the attosecond pulse techniques opened the door for studying and eventually controlling the electronic motion. As long as the notion of electronic states is valid, the potential-energy landscape in which the nuclei move is formed by the faster moving electrons. It is, therefore, very appealing to use attosecond pulses that modulate the electronic motion such that it triggers some desired nuclear rearrangement. Using the electron dynamics and the quantum coherence to induce a particular chemical process is the new paradigm of the 
emerging field of ``attochemistry'' (see, e.g., Ref.~\cite{NatPhot14_comm,FrMiMa_NatPhot14}).


This novel concept was to a large extent motivated by the interpretation of the results of a series of pioneering experiments performed by Schlag, Weinkauf, and co-workers (see, e.g., Refs. \cite{Weinkauf_1,Weinkauf_2,Schlag_1}) showing a charge-directed reactivity \cite{CDReact_CPL98} in electronically excited ionic states of various peptide chains. It was observed that after localized ionization of the chromophore site of the peptide, the positive charge is transferred to the remote end of the chain causing a bond-breaking. Time-resolved measurement on a smaller, prototype molecule (2-phenylethyl-\textit{N},\textit{N}-dimethylamine, abbreviated as PENNA) showed that the process takes place on the time scale of few tens of femtoseconds \cite{Lehr2005,Cheng2005}. Extensive \textit{ab initio} many-body calculations on PENNA suggested \cite{penna_CPL08} as explanation that the charge-directed reactivity is concerted electron-nuclear dynamics: immediately after the ionization pure electron dynamics are triggered and the positive charge starts to oscillate between the chromophore and the remote amine end of the molecule on a few-femtosecond time scale, while at later times the coupling to the slower nuclear dynamics causes the trapping of the charge on the amine site and the bond breaking. 

How is the charge transferred in this pure electronic step? In their pioneering work \cite{CM_first}, Cederbaum and Zobeley demonstrated that such a transfer, termed \textit{charge migration}, can be solely driven by the many-body effects (electron correlation and electron relaxation). Due to the electron correlation, the removal of an electron from a molecular orbital will create an electronic wave packet (a simultaneous population of a multitude of cationic states) which will evolve in time. The charge migration has been intensively studied theoretically \cite{Breidbach03,RelaxSat,jumping_JPCA10,RemLev_PRA12,Vitali_PRL13} and turned out to be a rich phenomenon, with many facets that are rather characteristic of the molecule studied (for a recent review, see Ref.~\cite{CM_review}). We note also that very recently, pure electronic, few-femtosecond charge oscillations were experimentally observed after a broadband ionization of phenylalanine \cite{Calegari2014} (see also Ref.~\cite{Greenwood_JPCL12}).


% ...

In this Letter we present a scheme for controlling the many-body electron dynamics of the charge migration process by a single ultrashort laser pulse. We show that by appropriately tailored infrared pulses one can stop the pure electronic, few-femtosecond oscillation of the charge, localizing it on the desired site of the molecule 3-methylen-4-penten-\textit{N},\textit{N}-dimethylamine (MePeNNA), which is a structural analogue of the mentioned above PENNA.

As noted already, the reason for the charge migration is that due to the electronic correlation, the removal of an electron from a particular molecular orbital populates several ionic states, creating in that way an electronic wave packet \cite{CM_first}. Depending on the type of the populated cationic states, different mechanisms of charge migration have been identified \cite{Breidbach03,RelaxSat,CM_review}, with the most common in the outer valence being the so-called hole-mixing mechanism. In the hole mixing, two (or more) ionic states appear to be linear combinations of two (or more) one-hole (1h) configurations, representing removal of an electron from a particular molecular orbital.

\begin{figure}[ht]
\begin{center}
\includegraphics[width=7.5cm]{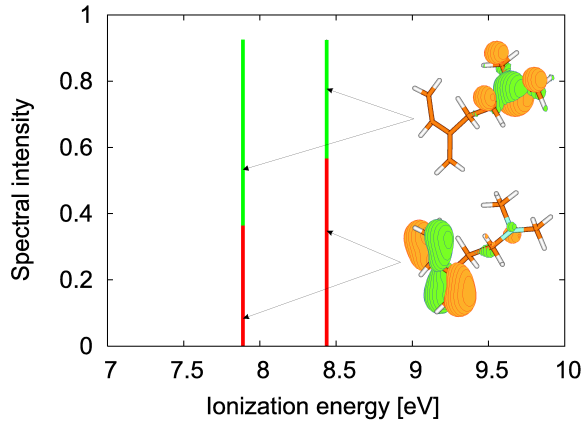}
\end{center}
\caption{\label{fig1}(color online) Ground and first excited cationic states of the molecule MePeNNA computed using the \textit{ab initio} many-body Green's function ADC(3) method~\cite{Schirmer1998}. The next ionic state is located at 10.5~eV. The contributions of the 1h configuration (HOMO)$^{-1}$ are given in red, while that of (HOMO$-1)^{-1}$ is shown in green. The two molecular orbitals are also depicted. The figure is adapted from Refs.~\cite{Luennemann_JCP08}.}
\end{figure}

A show-case example of a strong hole mixing appears in the molecule MePeNNA. Our many-body calculations showed that due to the electron correlation the ground and the first excited ionic states of the molecule are a strong mixture of two 1h configurations: an electron missing in the highest occupied molecular orbital (HOMO) and an electron missing in the HOMO$-1$, see Fig.~\ref{fig1} and Refs.~\cite{Luennemann_JCP08,emission}. Therefore, if we suddenly remove an electron either from HOMO or from HOMO$-1$, we will create an electronic wave packet, which in the Schr\"{o}dinger representation reads (in atomic units):
\begin{equation}\label{wf}
	|\Phi_i(t)\rangle = c_1(t) e^{-i E_I t}|I\rangle + c_2(t) e^{-i E_J t} |J\rangle.
\end{equation}
In the above expression $c_1(t)$ and $c_2(t)$ are the time-dependent (in general complex) amplitudes of the cationic eigenstates $|I\rangle$ and $|J\rangle$, satisfying $|c_1(t)|^2+|c_2(t)|^2=1$ at all times. Due to the hole-mixing structure of the ionic states, the evolution of the system described by the the wave packet (\ref{wf}) will represent an oscillation of the hole charge between the two involved 1h configurations, or between HOMO and HOMO$-1$. Since HOMO is localized on the chromophore and HOMO$-1$ on the amine group of the molecule (see Fig.~\ref{fig1}), the charge migration will represent an oscillation of the charge between the two ends of the molecule with frequency determined by the energy difference of the two states $\omega_0 = E_J-E_I$, which is 0.55~eV, meaning that the time needed for the charge to reach the remote end of the molecule is only 3.8~fs \cite{Luennemann_JCP08,emission}.

Interestingly, the ionization spectrum of MePeNNA suggests that an initial state of the form of Eq.~(\ref{wf}) can be achieved experimentally without approaching the sudden-approximation limit of removing an electron from a single orbital. As we see from Fig.~\ref{fig1}, for preparing such a wave packet we need a coherent population of the first two ionic states only, which can be done via a laser pulse with photon energy centered between the two states and a band width sufficient to embrace both of them. Since the states are about 0.5~eV apart, one needs a pulse with duration of about 1~fs. The next ionic state is located at 10.5~eV and therefore its population by such a pulse will be negligible. The initial localization site of the charge (chromophore or N-terminal) is determined by the relative phase between the two ionization channels. The latter can be controlled through the ionization pulse parameters, e.g., by chirping the pulse \cite{Cao_PRL98}, or by using a $\pi$-pulse \cite{pipulse_JCP00}.

We now pose the question whether after the electron dynamics are triggered and the charge starts to bounce back and forth between the two ends of the system, we can control its motion by applying a short laser pulse. The interaction of the  system with an external electric field $\vec{E}(t)$ can be described (in the dipole approximation) by the Hamiltonian
\begin{equation}\label{htot}
	\hat{H}(t)=\hat{H}_{ion} - \vec{D} \cdot \vec{E}(t),
\end{equation}
where $\hat{H}_{ion}$ is the full electronic Hamiltonian of the ionized system and $\vec{D}=(\hat{D}_x,\hat{D}_y,\hat{D}_z)$ is the vector operator of the dipole moment. The evolution of the system is governed by the time-dependent Schr\"odinger equation with the formal solution
\begin{equation}\label{prop}
	|\Phi_i(t)\rangle=\mathcal{T}\exp\left(-i \int_0^t \hat H(t')dt' \right)|\Phi_i(0)\rangle.
\end{equation}
In practice, one usually breaks the interval of interest $[0,t]$ into a large number of small increments, or time steps $\Delta t$, and Eq.~(\ref{prop}) is integrated numerically.

To describe the correlated motion of 69 electrons (the number of electrons in MePeNNA cation) in the presence of a laser field is an extremely difficult problem. To solve it we use \textit{ab initio} methods only. The cationic Hamiltonian $\hat{H}_{ion}$ is constructed using the so-called non-Dyson algebraic diagrammatic construction (ADC) scheme \cite{Schirmer1998} for representing the one-particle Green's function. At the ADC(3) level, used in the present calculations, the Hamiltonian is consistent with the exact Green's function up to the 3rd order of perturbation theory with respect to the Hartree-Fock (HF) reference Hamiltonian. Standard DZ basis sets \cite{Dunning} were used for constructing the non-correlated reference states.

One can include the interaction of the field by diagonalizing the cationic Hamiltonian matrix and using the field-free eigenstates as a basis to expand the dipole operator, computing the transition dipole matrix elements. Alternatively, when the diagonalization is very expensive, one can represent the dipole operator $\vec{D}$ in the many-electron basis of the molecular Hamiltonian (in the present case this is the so-called intermediate-state-representation basis \cite{Trofimov05}) and then directly propagate with the full Hamiltonian \cite{dprop}. For this purpose, or for performing the wave-packet propagation, Eq.~(\ref{prop}), we used the short-iterative Lanczos technique \cite{Leforestier}. Details of this technique allowing to study correlated-electron dynamics in systems containing few 10s of electrons are given in Ref.~\cite{dprop}.

Our aim is to find laser fields $\vec{E}(t)$ which can steer the evolution of the wave function of a system. As mentioned above, under certain ionization conditions, the initial state of the MePeNNA molecule will be a linear combination of the two lowest eigenstates of the ionic Hamiltonian and, therefore, can be regarded as a two-level system. There are numerous protocols suggested in the literature for controlling quantum dynamics in two-level systems (see, e.g., Refs. \cite{BrumShap_CPL86,Tannor_JCP86,STIRAP_RMP98,Barnes_PRA13}). Most of these protocols, however, are designed to perform a population inversion, that is, to invert the population distribution between the two states. This is insufficient for our purposes, as we would like to have the freedom to choose the desired final distribution of the populations starting from any initial one.

Very recently, we proposed a general approach for obtaining resonant laser pulses that can drive the populations of a two-level system in a predefined way \cite{Golubev2014}. If we want that the evolution of the system follows a particular quantum path, that is, that one of the populations in Eq.~(\ref{wf}) evolves according to some control function $f(t)$, i.e., $|c_1(t)|^2=f(t)$, then the field which can drive this transition takes the form \cite{Golubev2014}:
\begin{equation}\label{controlf}
	E(t)=\frac{1}{\mu} \frac{\dot{f}(t)}{\sqrt{f(t)(1-f(t))}} \sin{(\omega_0 t + \varphi)},
\end{equation}
where $\mu$ is the projection of the dipole moment on the electric field polarization axis, and $\varphi$ is the relative phase between the initial amplitudes, $c_1(0)$ and $c_2(0)$, of the populated ionic states. A convenient choice for the control function is \cite{Golubev2014}
\begin{equation}\label{ex_f}
	f(t)=a_{i} \left(1-\frac{1}{1+e^{-\alpha t}}\right)+a_{f} \frac{1}{1+e^{-\alpha t}},
\end{equation}
where $a_{i}$ and $a_{f}$ are the initial and the final (target) populations, respectively, of one of the states and the parameter $\alpha$ connects the transition time with the intensity of the field.

It should be noted that the two-level model is used here only for obtaining the field parameters. For computing the evolution of the hole charge in MePeNNA molecule in the presence of the control pulse, the propagation was performed with the full Hamiltonian, Eq.~(\ref{htot}), as described above.

A convenient quantity for visualization, or for tracing in time and space the charge dynamics is the so-called hole density \cite{CM_first,Breidbach03}. The hole density is defined as the difference between the electronic density of the neutral and that of the cation
%
\begin{equation}\label{hole_dens}
	Q(\vec r,t) = \langle\Psi_0|\hat\rho(\vec r)|\Psi_0\rangle -
	\langle\Phi_i(t)|\hat\rho(\vec r)|\Phi_i(t)\rangle,
\end{equation}
where $\hat{\rho}(\vec r)$ is the one-body electronic density operator, $|\Psi_0\rangle$ is the ground state of the neutral, and $|\Phi_i(t)\rangle$ is the propagated cationic wave packet. The quantity $Q(\vec r,t)$ describes the density of the hole at position $\vec r$ and time $t$ and by construction is normalized at all times~$t$.

Let us now examine two situations of particular interest for achieving control over the charge migration dynamics, namely stopping the charge oscillations and localizing the charge on one of the two molecular sites. To be specific, we will assume that the initial ionization of our test molecule, MePeNNA, is performed such that the electron is removed from the HOMO, i.e. the initial hole charge is localized on the chromophore. As discussed above, this will trigger pure electron dynamics in which the charge will oscillate between the chromophore and the amine site and we would like to apply a control pulse which will stop this oscillation and localize the charge on the amine group or on the chromophore.

Such a control can be achieved by a laser pulse obtained via Eqs.~(\ref{controlf}) and (\ref{ex_f}) by choosing the desired initial and final populations of the two states in the wave packet. 
In the case of MePeNNA, if we want to drive the system to a stationary state in which the charge is entirely localized in the HOMO, we need to choose $a_{f}=1$, while if we want to localize the charge on the amine site, we need to take $a_{f}=0$. The initial population is $a_{i}=0.4$, reflecting the fact that an initial state with a hole localized in the HOMO has the form $|\Phi_i(0)\rangle=\sqrt{0.4}|I\rangle + \sqrt{0.6}|J\rangle$ (see also Fig.~\ref{fig1}).

As mentioned above, the parameter $\alpha$ in Eq.~(\ref{ex_f}) controls the interplay between the duration and the intensity of the pulse needed to perform the transition~-- a slow transition can be achieved with a weak pulse, while shorter pulse will naturally need higher intensity. Since we want to modulate the charge migration before the nuclear motion will start to influence the dynamics, we would like to use as short pulses as possible. On the other hand, the high intensity of the control pulse may lead to undesired multiphoton processes, which can, for example, further ionize the system. Therefore, we need to balance between these two factors. In the case of MePeNNA the minimum pulse duration for performing the needed transition is about 10~fs, giving an electric field strength which never exceeds 10$^9$~V/m. This corresponds to a pulse with a peak intensity of about 10$^{11}$~W/cm$^2$, which is rather weak.

\begin{figure}[ht]
\begin{center}
\includegraphics[width=7.8cm]{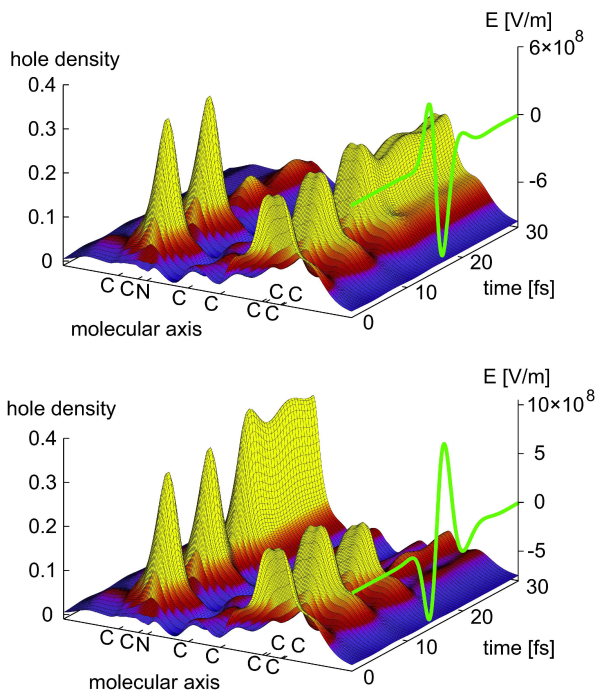}
\end{center}
\caption{\label{fig2}(color online) Time evolution of the hole density, Eq.~(\ref{hole_dens}), along the molecular axis of the molecule MePeNNA, after an initial localized ionization of the chromophore, controlled with a laser pulse (shown on the right) centered at 15~fs. The molecular axis is chosen to pass through the longest spatial extension of the molecule. Upper panel: the laser pulse is designed to achieve localization of the charge on the chromophore. Lower panel: the laser pulse is designed to achieve localization of the charge on the amine site.}
\end{figure}

The results of our full propagation accounting for the influence of the control field with the above parameters are shown in Fig.~\ref{fig2}. The figure depicts the hole density, defined in Eq.~(\ref{hole_dens}), of the molecule MePeNNA after creating the initial hole on the chromophore at time $t=0$ and applying a control laser pulse (also shown in the figure) with the maximum of the field centered at $t=15$~fs. Immediately after the ionization, the charge migration dynamics take place with the hole jumping from one end of the system to the other. We clearly see that between $t\sim 10$~fs and $t\sim 20$~fs, the time during which the system is exposed to the control pulse, the charge oscillations are nearly completely stopped and the hole becomes localized on the desired site~-- on the chromophore (upper panel of Fig.~\ref{fig2}) or on the amine group (lower panel of Fig.~\ref{fig1}). We would like also to emphasizes that the charge stays put at the desired site of the molecule \textit{after} the pulse is over, that is, the pulse is tailored such that it brings the system to a superposition of electronic states in which the density is essentially stationary. The small oscillations taking place after the pulse is over reflect the fact that the molecule is, of course, not a perfect two-level system. Because of the electron correlation, in addition to the two mixed 1h configurations, the two involved ionic eigenstates contain also small contributions from the two-hole--one-particle (2h1p) configurations \cite{breakdown86}. The latter represent excitations on top of the ionization and their weight forms the missing to 1 part in the states depicted in Fig.~\ref{fig1}. Through the 2h1p configurations, other molecular orbitals also contribute to the dynamics and get population, while the pulse is optimized to account only for the HOMO and the HOMO$-1$. Nevertheless, the suggested simple scheme for obtaining the needed control-pulse parameters works remarkably well in such a complicated system as the MePeNNA molecule.

Let us now comment on the possible further evolution of the studied system. Due to its similarity with the mentioned above PENNA molecule, one may expect a similar charge-directed reactivity after localized ionization of the chromophore. The charge starts to oscillate between the two ends of the molecule until the nuclear motion eventually traps it at the amine site and the molecule dissociates by braking the bridging carbon-carbon bond. However, as we noted above, the nuclear motion is strongly influenced by the electron dynamics and, therefore, by controlling the charge migration we may be able to predetermine the nuclear rearrangement. Localizing the charge on the chromophore may substantially slow down or even prevent the dissociation, while its localization on the amine site will most probably speed-up the break-up of the molecule.

To check this plausible hypothesis, one needs a full quantum treatment of the coupled electron-nuclear dynamics. This is unfortunately currently out of reach for so large systems. We would like to note, however, that a semiclassical method that might give some hints about how the nuclear motion will be affected by the electron dynamics, was recently proposed \cite{MendiveTapia2013,Vacher2014,Vacher_JCP14} and could be a good starting point for such studies.

Before concluding, we would like to emphasize that the scheme presented in this work is general and not restricted to only stopping the charge migration oscillations. Through the control function $f(t)$ in Eq.~(\ref{controlf}), one is able to obtain the pulse shape needed to drive the system to any combination of the two electronic states, and thus bring the ion to the optimum initial condition for the desired follow-up nuclear motion. 

We hope that our study will stimulate further theoretical and experimental work on the possibilities to control chemical reactions via the manipulation of the electron dynamics.

The authors thank Lorenz Cederbaum for many valuable discussions and Markus Pernpointner for the technical help in performing the calculations. Financial support by the DFG and by the US ARO (grant number W911NF-14-1-0383) is gratefully acknowledged.

\end{document}